\documentstyle[12pt,epsf]{article}
\begin{document}
\title{Dyon-Skyrmion Lumps}

\author{{\large Y. Brihaye}$^{\diamond}$,
B. Kleihaus$^{\dagger}$
and {\large D. H. Tchrakian}$^{\dagger \star}$ \\ \\
$^{\diamond}${\small Physique-Mathematique, Universite de Mons-
Hainaut, Mons, Belgium}\\ \\
$^{\dagger}${\small Department of
Mathematical Physics, National University of Ireland Maynooth,} \\
{\small Maynooth, Ireland} \\ \\
$^{\star}${\small School of Theoretical Physics -- DIAS, 10 Burlington Road,
Dublin 4, Ireland }}

\date{}
\newcommand{\dd}{\mbox{d}}\newcommand{\tr}{\mbox{tr}}
\newcommand{\ee}{\end{equation}}
\newcommand{\be}{\begin{equation}}
\newcommand{\ii}{\mbox{i}}\newcommand{\e}{\mbox{e}}
\newcommand{\pa}{\partial}\newcommand{\Om}{\Omega}
\newcommand{\vep}{\varepsilon}
\newcommand{\bfph}{{\bf \phi}}
\newcommand{\lm}{\lambda}
\def\theequation{\arabic{equation}}
\renewcommand{\thefootnote}{\fnsymbol{footnote}}
\newcommand{\re}[1]{(\ref{#1})}
\newcommand{\bfR}{{\sf R\hspace*{-0.9ex}\rule{0.15ex}%
{1.5ex}\hspace*{0.9ex}}}
\newcommand{\N}{{\sf N\hspace*{-1.0ex}\rule{0.15ex}%
{1.3ex}\hspace*{1.0ex}}}
\newcommand{\Q}{{\sf Q\hspace*{-1.1ex}\rule{0.15ex}%
{1.5ex}\hspace*{1.1ex}}}
\newcommand{\C}{{\sf C\hspace*{-0.9ex}\rule{0.15ex}%
{1.3ex}\hspace*{0.9ex}}}
\renewcommand{\thefootnote}{\arabic{footnote}}

\maketitle
\begin{abstract}
We make a numerical study of the classical solutions of the combined system
consisting of the Georgi-Glashow model and the $SO(3)$ gauged Skyrme model.
Both monopole-Skyrmion and dyon-Skyrmion solutions are found. A new
bifurcation is shown to occur in the gauged Skyrmion solution sector.
\end{abstract}
\medskip
\medskip
\newpage

\section{Introduction}

There are two 3+1 dimensional $SO(3)$ gauge field models which support
static soliton solutions. One is the Georgi-Glashow (GG) model which
supports the well known monopole \cite{tp}, and the other is the
$SU(2)_{L+R}$, or vector,
gauged Skyrme model \cite{AT,BT} which also supports $SO(3)$ gauged
Skyrmions. In addition to the monopole, the GG model supports also dyon
solutions \cite{jz} which in addition to the magnetic charge carry an
electric charge as well.
The topological stability of the monopole comes from the magnetic charge,
which is descended from the second Chern-Pontryagin charge, while the
topological charge of the gauged Skyrmion is the degree of the map.

Combining these two
models, we have a new system whose topological charge is a sum of the
respective charges, and it can reasonably be expected that this system also
supports static finite energy solitons. Note that in this case the local
$SO(3)$ symmetry is broken down to $U(1)$ via the Higgs mechanism, in
contrast to the $SO(3)$ gauged Skyrme model on its own, in which case the
local $SO(3)$ symmetry is not broken at all and 3 massless gauge bosons
survive. In this preliminary investigation, this is precisely what we have
done. Using numerical methods, we verify that such solutions exist.
Moreover, we have sought and found both monopole-Skyrmion and dyon-Skyrmion
solutions, and studied some of their properties. The combined system
supports solutions also
with zero monopole charge, unit Baryon charge, as well as with unit
monopole charge, zero Baryon charge.

Even though this is a self contained numerical study of the classical
solutions alluded to above, it is in order to put it into context both in
the background of previous work involving the gauging of the Skyrme model
\cite{S}, and, from the viewpoint of its potential physical relevance.

The Skyrme model was gauged by Witten in Ref.~\cite{W}, and others e.~g. in
Refs. \cite{BPR}. These works were carried out in the context of current
algebra results, and were not concerned with the solitonic aspects of the
gauged Skyrmion. That was done subsequently by many authors \cite{C} where
gauged Skyrme solitons were studied with the aim of explaining the low
energy properties of Hadrons. Also in the context of electroweak theory,
which can be regarded as a gauged Skyrme model in the limit of
very high Higgs mass, Rubakov \cite{R1} and Eilam {\it et al.}~\cite{EKS}
considered the static classical
solutions of the $SU_L$ gauged Skyrme model. In all these cases, there is
no topological lower bound and the classical solutions are metastable, but
for certain values of the parameters in one of these models \cite{R1,EKS} a
stable branch
of solitons appears as a result of catastrophic behaviour. The advantage of
the gauging used in Refs. \cite{R1,EKS} is that the 4-divergence of the
topological current does not vanish but equals the local chiral anomaly
\cite{GW,DF}, which can present itself as a mechanism for Baryon number
violation as explained in Ref.~\cite{R1}.

In the context of Baryon number violation, there is an older mechanism
suggested by Rubakov \cite{R2} and by Callan \cite{C} where
monopole-(left-handed massless)
Fermion interactions lead to Fermion number nonconservation. The mechanism
involves the fluctuations of the electric field, in the presence of the
magnetic field of the monopole, giving rise to nonzero chiral anomaly and
hence Fermion number violation. This was shown for the case of massless
(left-handed) Fermions,
by scattering with the monopole, which describes a high energy process. The
approximation techniques employed \cite{C,R2} are neither perturbation
theoretic
nor semiclassical. To describe a low energy process such as a decay, it
would be more appropriate to deal with a process that is susceptible to
semiclassical analysis. To this end, Callan and Witten \cite{CW} replaced
the massless Fermions by the Skyrme soliton \cite{S}, interacting with
the (Abelian) magnetic field of the monopole. While they \cite{CW} did not
seek to demostrate the existence of a $U(1)$ gauged Skyrmion, this is
implicit in their work and has recently been verified numerically
\cite{PT}. In the background of this it is hoped that the present work,
which sets out to find the monopole-Skyrmion and dyon-Skyrmion solutions,
would be of concrete usefulness to a semiclassical method of describing
Baryon number decay. In particular the dyon-Skyrmion excites a nonzero
classical quantity for the chiral anomaly, which can lead to chirality
breaking as pointed out long ago by Marciano and Pagels \cite{MP}.

In Section 2 we present the model and give the topological lower bounds
 on the
static energy. In Section 3 we give the static spherically symmetric fields
and the field equations in the static limit. Sections 4 and 5 deal,
respectively, with the results of the numerical analysis of the 
${\bf A}_0 =0$
and ${\bf A}_0 \neq 0$ cases. Section 5 in particular, includes an in depth
analysis of the Julia-Zee dyon \cite{jz}. We summarise and discuss our
results in Section 6.

\section{The Model}

The model under consideration is the combination of the Georgi-Glashow (GG)
model and of the $SO(3)$ gauged $O(4)$ (Skyrme) model studied previously in
Refs.
\cite{AT,BT}. We state the Lagrangian of each of these models separately,
defined in four dimensional Minkowski space, each being normalised properly
so that the value of the energy of the static soliton in each case lies
above its own
topological lower bound. The static solutions in question satisfy the
Euler- Lagrange equations of the static energy density functional, which is
the static hamiltonian in the temporal gauge. In the GG case, this is the
'tHooft- Polyakov \cite{tp} monopole, while in the latter case it is the
soliton studied in Refs. \cite{AT,BT}.

The GG model is described by
\be
\label{2.1}
{\cal L}_{GG}=
-{1\over 4}  \lambda_0^4 |F_{\mu \nu}^{\alpha}|^2
 + {1\over 2}  \lambda_1^4
|D_{\mu} \Phi^{\alpha}|^2 -{1\over 4} \lambda_2^4 (\eta^2 -
|\Phi^{\alpha}|^2)^2 , \ee
\be
\label{2.2}
F_{\mu \nu}^{\alpha} =\pa_{\mu} A_{\nu}^{\alpha} -\pa_{\nu}
A_{\mu}^{\alpha} +\vep^{\alpha \beta \gamma} A_{\mu}^{\beta}
A_{\nu}^{\gamma},\: \: \: D_{\mu} \Phi^{\alpha}=\pa_{\mu} \Phi^{\alpha}
+\vep^{\alpha \beta \gamma} A_{\mu}^{\beta} \Phi^{\gamma} . \ee The late
Greek indices $\mu ,\nu ,..$ label the Minkowski space vectors, while the
early Greek indices $\alpha ,\beta, ..=1,2,3$ label the elements of the
algebra of the gauge group $SO(3)$. The Latin letters $a,b,..=1,2,3,4$, so
that $a=(\alpha ,4)$ are reserved for the $O(4)$ Skyrme model. 
In eq.~\re{2.1}
the constant $\eta$ is the  vacuum expectation value 
(VEV) of the Higgs field and like the latter has
the inverse dimension of a length. 
 The constants $\lambda_0 ,\lambda_1$ and
$\lambda_2$ are all dimensionless.

The $SO(3)$ gauged Skyrme model is described by \be \label{2.3} {\cal
L}_{O(4)} =
-{1\over 4} \kappa_0^4 |F_{\mu \nu}^{\alpha}|^2 +{1\over 2} \kappa_1^2
|D_{\mu} \phi^a|^2 -{1\over 8} \kappa_2^4 |D_{[\mu} \phi^a D_{\nu
]}\phi^b|^2 \ee
\be
\label{2.4}
D_{\mu} \phi^{\alpha}=\pa_{\mu} \phi^{\alpha} +\vep^{\alpha \beta \gamma}
A_{\mu}^{\beta} \phi^{\gamma} , \qquad D_{\mu} \phi^4 =\pa_{\mu} \phi^4 .
\ee
 The constants $\kappa_0$ and $\kappa_2$ are dimensionless and 
the constant $\kappa_1$ has dimension 
of a inverse length.
The reason we keep all the coupling constants arbitrary in eqs.~\re{2.1} 
and \re{2.3} will appear soon.

When we consider the static Hamiltonians corresponding to the Lagrangians
above in the temporal gauge, i.~e. $A_0 = 0$, we can write the following
topological identities: 
\be
\label{topgg}
\int d {\bf r} \ {\cal H}_{GG} \ge 4 \pi \eta  \lambda_0^2 \lambda_1^2
 M \ , \ee \be
{\cal H}_{GG} =
{1\over 4}  \lambda_0^4 |F_{j k}^{\alpha}|^2 
+{1\over 2}  \lambda_1^4 |D_{j}
\Phi^{\alpha}|^2 +{1\over 4} \lambda_2^4 (\eta^2 - |\Phi^{\alpha}|^2)^2 , 
\ee
where the integer $M$,
representing the index of the mapping $\Phi^{\alpha}(\vec x)$, is the
monopole topological charge. Similarly \cite{BT} \be \label{topsk} \int d
{\bf r} \ {\cal H}_{O(4)} \ge
12 \pi^2 \kappa_1 \kappa_2^2
{1 \over {2 \sqrt{1 + 9 ({\kappa_2 \over \kappa_0})^4}}} T \ee \be {\cal
H}_{O(4)} =
{1\over 4} \kappa_0^4 |F_{j k}^{\alpha}|^2 +{1\over 2} \kappa_1^2 |D_{j}
\phi^a|^2
+{1\over 8} \kappa_2^4 |D_{[j} \phi^a D_{k ]}\phi^b|^2 \ee where the
integer $T$, representing the index of the mapping $\phi^a(\vec x)$, is the
Skyrmion topological charge. In the Skyrme description of hadrons, $T$ is
identified with the baryon number.

In the following will consider also the equations resulting from the
superposition of the two lagrangians eqs.~\re{2.1} and 
\re{2.3} \be \label{2.5}
{\cal L}_m ={\cal L}_{GG} +{\cal L}_{O(4)} . \ee The finite energy
configurations of this mixed lagrangian are characterized by the couple of
integers $M,T$. Classical solutions corresponding to the two different
topological excitations can then be constructed, they correspond to the
configuration with minimal energy in a class $M,T$.

In order to normalize the fields conventionally, we have to choose \be
 \lambda_0^2 = {1\over e} \cos(\theta) \ \ , \ \ \kappa_0^2 = {1\over e}
\sin(\theta) \ \ , \ \  \lambda_1^4 = 1
\ee
where $e$ denotes the gauge coupling constant. With the choice $\theta =
\pi/4$, the topological inequality relating ${\cal H}_m$ to the class of
solutions of indexes $M,T$
reads
\be
\label{topmix}
\int d {\bf r} \ {\cal H}_m \ge
{4 \pi \eta \over e}
( {1\over \sqrt{2}} M + {3 \pi \over 2}
{\sqrt{\xi \kappa \over \sqrt{1 + 18 \kappa}}} T ) \ee where
\be
\label{lam}
\lambda = { \lambda_2^4 \over e^2} \ \ , \ \ \xi = {1 \over \eta^2} 
\kappa_1^2
\ \ , \ \ \kappa = e^2 \kappa_2^4 .
\ee
In eq.~\re{lam}, $\lambda$ is defined for later convenience. Note that the
topological lower bound eq.~(\ref{topmix}) can be refined by an optimal 
value
of the mixing angle $\theta$, depending on the parameters 
$\lambda_1 ,\lambda_2$,
$\kappa_1 , \kappa_2$. To acheive this it is necessary to solve a
complicated non linear equation, which we shall not pursue here.

\section{Static spherically symmetric equations}

The classical equations corresponding to 
eqs.~\re{2.1}, \re{2.3} and \re{2.5} 
are in general intractable.
We will restrict our search of solutions to the static and spherically
symmetric case.
If we choose to employ the temporal gauge in the static limit, the Euler-
Lagrange equations will reduce to the variational equations arising from
the static Hamiltonians pertaining to the Lagrangians eqs.~\re{2.1} and
\re{2.3}. The latter would be bounded from below by the monopole charge 
and
the Baryon number densities, respectively. Hence the solutions to the
classical equations of each of these static Hamiltionians, separately, can
describe the 'tHooft-Polyakov monopole \cite{tp} and the soliton of the
$SO(3)$ gauged Skyrme model \cite{AT,BT}. The Euler-Lagrange equations of
the Hamiltonian of the combined static system, i.~e. GG-Skyrme, in the
temporal gauge also supports soliton solutions since the Hamiltonian is
again bounded from below by the two topological charges eq.~(\ref{topmix}).
This is one of the problems
studied in the present work yielding the monopole-Skyrmion solitons.

If instead of employing the temporal gauge we proceed like Julia and Zee
\cite{jz} and solve the Euler-Lagrange equations pertaining to the
Lagrangian eq.~\re{2.5} defined on Minkowski space in the static limit, the
resulting solutions of the GG-Skyrme system describe the dyon-Skyrmion.
This is the other problem studied in this work. As in the case of the dyon
\cite{jz} on its own, we shall restrict to the spherically symmetric
solutions only \footnote{In this case the classical equations simplify
sufficiently to become tractable. To our knowledge the only dyon solutions
known are the spherically symmetric Julia-Zee\cite{jz} dyons.}.

The spherically symmetric Ansatz employed is \be \label{3.1} A_i^{\alpha}
=\frac{a(r)-1}{r} \vep_{i\alpha \beta} \hat x^{\beta}, \qquad A_0^{\alpha}
= \frac{g(r)}{r}\: \hat x^{\alpha} \ee \be \label{3.2}
\Phi^{\alpha} =\eta \: h(r)\: \hat x^{\alpha} \ee \be \label{3.3}
\phi^{\alpha} =\sin f(r) \: \hat x^{\alpha} ,\qquad \phi^4 =\cos f(r) . \ee
Notice that the functions $a(r)$, $h(r)$, $g(r)$ and $f(r)$ are
dimensionless. We
find it useful to introduce a dimensionless radial variable \be x = M_W r \
\ \ , \ \ \ M_W \equiv e \eta \ee

Substituting the Ans\"atze eqs.~\re{3.1}-\re{3.3} into the static limit
 of the
Lagrangian eq.~\re{2.5}, leads to the following one
dimensional (radial) Lagrangian density $L_m$, defined by \be
\label{ec}
\int L_m \: dx = \int {\cal L}_m \: d {\bf r} = E_1 - E_2 \ee with \be
E_p \equiv {4 \pi \over e} \eta \tilde E_p = {4 \pi \over e} \eta \int dx
{\cal E}_p
\ \ , \ \ p=1,2
\ee
\begin{eqnarray}
{\cal E}_1
&=& {1 \over 2} x^2 (g')^2 + a^2 g^2 \label{ed1} \\ {\cal E}_2 &=& (a')^2 +
{(a^2- 1)^2\over{2x^2}} \nonumber \\ &+& {1\over 2} x^2(h')^2 + a^2 h^2 +
{\lambda\over 4} x^2(h^2-1)^2 \nonumber \\ &+& {\xi \over 2} [x^2(f')^2 + 2
a^2 \sin^2f] \nonumber \\ &+& \kappa a^2 \sin^2f[(f')^2 + a^2{\sin^2f\over
{2 x^2}}] \label{ed2} \end{eqnarray} where we have separated the
contribution $E_1$ due to the electric field and the prime denotes the
derivative with respect to $x$.

The static classical equations corresponding to the Lagrangian density
${\cal L}_m$, in the spherically symmetric Ansatz, turn out to be
equivalent to the
equations obtained by varying the effective one dimensional density (see
eqs.~\re{ed1} and \re{ed2}) ${\cal E}_1-{\cal E}_2$ with respect to 
the radial
functions $a,g,h$ and $f$. These equations are obtained straightforwardy
and we do not list them here. 
We note however, that for each function the corresponding variational
equation can be solved trivially by setting this function to zero.

%We note however that each of these is
%trivially solved when the function it describes the variation of, vanishes.

It will be useful to present their asymptotic forms in the $x\gg 1$ region,
to facilitate subsequent explanations. They are, in order of the variations
of $a,g,h$ and $f$~:
\begin{eqnarray}
\label{eqa}
&a'' &= a({a^2-1 \over x^2} + h^2 - g^2 + \xi \sin^2 f + \dots) \\
\label{eqc} &(x^2 g')' &= 2 g\: a^2 \\
\label{eqh}
&(x^2 h')' &=h(2 a^2 + \lambda x^2 (h^2-1)) \\ \label{eqf} &(x^2 f')' &= 2
a^2 \sin f \cos f + o(\kappa / \xi) . \end{eqnarray}

Following \cite{jz}, we define the energy of a configuration by $E = E_1 +
E_2$, which coincides with the volume integral of the static Hamiltonian
obtained in the usual way from the gauge invariant stress tensor. The
topological lower bound for $E_2$
follows immediately from eqs.~(\ref{topgg}), (\ref{topsk}) and
 (\ref{topmix}).

\section{Numerical results, case $ {\bf A}_0 =0$}

We first discuss the classical solutions in absence of the electric field,
i.~e. with $g(x)=0$.
Eq.~(\ref{eqc}) is trivially solved and we are left with a system of
three non linear differential equations. Only the part $E_2$ of the action
is relevant in this case. In the following, we will conveniently denote the
value $\tilde E_2$ of the solution with given $M$ and $T$ by \be
\label{emon}
E_{MT} (\lambda,\xi,\kappa) . \ee
We now describe the four cases with $M\leq 1$ and $T\leq 1$.

\subsection{Case $M=0,T=0$}
This corresponds to the class of the vacuum which is not spherically
symmetric. It has a zero energy \be
E_{00} (\lambda, \xi, \kappa)=0 \ . \ee

\subsection{Case $M=1,T=0$}
This case corresponds to the celebrated SU(2) magnetic monopole \cite{tp}.
Since $T=0$, it has $f(r)=0$; as a consequence, the parameters $\xi$ and
$\kappa$ are irrelevant for this case. The boundary conditions and
asymptotic behaviour of the functions $a,h$ read \begin{eqnarray}
\label{moor}
&a(0)=1\quad , \quad &h(0)=0 \\
\label{moas}
&a(x) \simeq A e^{-x}\quad , \quad &h(x)\simeq 1 -B e^{- \sqrt{2 \lambda
}x} \quad \ \ ( x\rightarrow \infty) \end{eqnarray} where $A,B,F$ are
constants.
The values of the energy of the monopole solution were computed long ago
\cite{bm}
(our numerics fully reproduces these values); the energy increases
monotonically with $\lambda$ as demonstrated in Table 1.

In the Bogomol'nyi limit, $\lambda = 0$, the energy coincides with the
topological lower bound,
i.~e. (omitting the parameters $\xi$ and $\kappa$) \be
E_{10}(\lambda) \geq E_{10}(0) = 1 \ \ . \ee The solution, the
Prasad-Sommerfield monopole, is expressed in terms of elementary functions
\cite{ps}. Its behaviour near the origin is given by (\ref{moor}) but, for
$x\rightarrow \infty$, we have \be a(x) \simeq x e^{-x}\quad , \quad
h(x)\simeq 1 - {1 \over x} \quad \ \ \ee instead of (\ref{moas}).

\subsection{Case $M=0,T=1$}

 In this case, there is no Higgs field and consequently no Higgs potential.
Correspondingly, the topological lower bound reads \be
E_{01}(\lambda, \xi, \kappa) \geq {3 \pi \over 2} {\sqrt{\xi \kappa} \over
\sqrt{1 + 9 \kappa}} \ee
independently of $\lambda$.
The static equations describe the gauged Skyrmion studied in 
Ref.~\cite{BT}.

The parameter $\xi$ can be changed by
a rescaling of the radial variable $x$.
Comparison of the energies of the gauged Skyrmion and of the monopole is
demonstrated in
Table 1 and Fig.~1. 
In Table 1 we list the energies for various values of $\lambda$ for 
fixed $\xi = 1$ and $\kappa = 0.4$,
while in Fig.~1 $\kappa$ varies and we have fixed $\lambda = \xi = 1$.

Let us now come to the detailed discussion of the solutions in the region
$\kappa \approx 0.8$. For completeness, it is useful to summarise the
possible boundary conditions available for the gauged Skyrmion. At the
origin $x=0$ the behaviour of the radial functions is uniquely determined
by the condition of continuity of the fields at the origin: \be
a(x) = 1 + A_1 x^2 + o(x^3) \ \ \ , \ \ \ f(x) = \pi + F_1 x + o(x^2) .
\label{gsor}
\ee
In contrast, in the $x\gg 1$ asymptotic region, several conditions are
consistent with the
finiteness of the energy. Classical solutions of the equations have been
obtained \cite{BT} with
the two following sets
\begin{eqnarray}
&{\rm Type \ A\ :} \ \ & a \approx 1 - {A \over x} \ \ \ , \ \ \ f \approx
{F \over x^2} \label{gsasa} \\ 
&{\rm Type \ B\ :} \ \ & a \approx {A \over
x^{\alpha}} \ \ \ , \ \ \ f \approx {F \over x}
\label{gsasb}
\end{eqnarray}
where $\alpha \equiv (\sqrt{4F^2 - 3} - 1)/2$.

The following results were obtained in \cite{BT}. For small values of
$\kappa$, the solution is of type $A$, its energy increases monotonically
from $E=0$ (for $\kappa = 0$) and the branch (say branch $A$) stops at a
critical value $\kappa = \kappa_A^{cr} \approx 0.8091$. For large values of
$\kappa$ (in fact for $\kappa > \kappa_B^{cr} \approx 0.69122$) 
the solution is
of type B. We call this branch B. By using arguments of catastrophe theory
\cite{EKS}, one can reasonably expect the occurence
of a third branch of solutions on the interval $\kappa \in [\kappa_B^{cr},
\kappa_A^{cr}]$, as was explained in \cite{BT}.

A third branch indeed
exists. The solutions on this branch obey the condition of type $A$ and
therefore we refer to it as
branch $\tilde A$. The energies of the three branches of solutions are
depicted on Fig.~2. The branches $A$ and $\tilde A$ terminate at $\kappa =
\kappa_A^{cr}$, forming a cusp catastrophy. The transitions of the 
profile of
the solutions from branch $A$ to branch $\tilde A$ is smooth.

In contrast, when the limit $\kappa \rightarrow \kappa_B^{cr}$ is 
considered,
the solutions of
the branch $\tilde A$ approach the limit of branch $B$ in a subtle way.
For instance, the value $x_m$ for which the function $a(x)$ has a minimum
(say $a_m$) tends to infinity, while $a_m$ tends to zero. For values of
$\kappa$ close to $\kappa_1$, the solutions of branches $\tilde A$ and $B$
coincide on a large interval of $x$ (typically on $x \in [0, 10^7]$ for
$\kappa = 0.6914$) and deviate from each other for larger values of $x$.
In the limit $\kappa \to \kappa_B^{cr}$ this interval becomes infinitely 
large
and the two solutions deviate at infinity. This can clearly be seen from
Fig.~3.
A similar demonstration can be made for the function $f(x)$, namely that 
near
the critical point $\kappa_B^{cr}$ these functions for the two solutions 
on brances
$B$ and $\tilde A$ also coincide. We do not display the graphs analogous
to Fig.~3 in this case. The behaviour of the solutions is further 
illustrated by
Figs.~4 and 5 where we plot respectively the value of $F_1$ (defined in
eq.~(\ref{gsor})) for the three branches and the value of $\alpha$ as a 
function
of $\kappa$.

Fig.~2 furnishes a simple interpretation of the three solutions. To discuss
it, we introduce $\kappa_{AB}^{cr}$ as the value of $\kappa$ where 
the energy of
the branches $A$ an $B$ coincide ($\kappa_{AB}^{cr} \approx 0.785$).
 On the
interval $\kappa \in [\kappa_B^{cr}, \kappa_{AB}^{cr}]$ the 
solution on branch $A$
constitutes the absolute minimum of the energy functional $E_2$, 
while the
one on branch $B$ is a local minimum. 
The solution on the branch $\tilde A$
is a sphaleron  corresponding to a saddle point which 
represents the energy barrier between the two minima. The
situation is similar on
the interval $\kappa \in [\kappa_{AB}^{cr}, \kappa_A^{cr}]$; 
the absolute (resp. local) minimum energy
configuration is then on branch $B$ (resp.$A$).
 As $\kappa$ approaches the critical value $\kappa_B^{cr}$ 
the local minimum of branch $B$ 
approaches the sattle point of branch $\tilde A$. At the critical $\kappa$  
both coincide and form an inflection point. For $\kappa<\kappa_B^{cr}$ 
this point is no longer an extremum and the solutions 
of branches $B$ and $\tilde A$ cease to exist.
The global minimum of branch $A$ is then the only extremum and only 
branch $A$ solutions exist. 
The same scenario applies at the other critical value $\kappa_A^{cr}$
where the solutions of branches $A$ and $\tilde A$ stop to exist and 
only branch $B$ solutions exist.

\subsection{Case $M=1,T=1$}

It is natural to call this solution the "monopole-Skyrmion". The three
functions $a,h,f$ are non trivial and obey the following boundary
conditions at $x=0$ and as $x\to \infty$, repectively, \begin{eqnarray}
\label{skmoas1}
a(0)=1\quad &, \quad h(0)=0\quad &, \quad f(0)=\pi \\ \label{skmoas2} a(x)
\simeq A e^{-x}\quad &, \quad h(x) \simeq 1- B e^{-\sqrt{2\lambda}x} \quad
&, \quad f(x) \simeq {F\over x}
\end{eqnarray}
where $A,B$ and $F$ are constants.
In contrast to the case of the gauged Skyrmion solution \cite{BT}, the
finite energy condition
leads to a unique asymptotic behaviour of the solution.

The energy of the solution is given in Table 1 for several values of 
$\lambda$ (for $\xi=1$ and $\kappa=0.4$), indicating that the
energy of the monopole-Skyrmion varies rather slightly with $\lambda$. 
The corresponding lower bound inequality reads
\be E_{11}(\lambda, \xi, \kappa) \geq {1 \over \sqrt{2}} + {3\pi \over 2}
{\sqrt{ \xi \kappa} \over \sqrt{1 + 18 \kappa}} .
\ee

\subsection{General properties}

We have constructed numerically the three non trivial topological
solitons above for numerous values of the coupling constants
$\lambda,\xi, \kappa$ and computed their energies. 
In order to
give an idea of the relative magnitudes for the different classes,
let us choose $\lambda=1, \xi=1, \kappa=0.4$, then
\be
E_{00} = 0\quad , \quad
E_{10} \approx 1.29\quad , \quad
E_{01} \approx 2.98\quad , \quad
E_{11} \approx 3.53 \ \ .
\ee
The Bogomol'nyi limit $\lambda=0$ is of particular interest since in that
case the monopole saturates its topological lower bound. Choosing again
$\xi=1, \kappa=0.4$, we find \be E_{00} = 0 \quad ,
 \quad E_{10} = 1.0 \quad ,
\quad
E_{01} \approx 2.98\quad , \quad
E_{11} \approx 3.45
\ee

The behaviour of the solutions in the limit $\kappa \rightarrow 0$,
with $\lambda, \xi$ fixed, was carefully analysed. Our numerical
analysis strongly supports the following formula~:
\be \lim_{\kappa\rightarrow 0}
E_{M 1} (\lambda, \xi, \kappa) = E_{M 0}(\lambda, \xi, 0) \ \ ,
\ \ {\rm for \ } M = 0,1 \ee as illustrated by Fig.~1. Indeed,
in the limit $\kappa \rightarrow 0$, the functions $a(r), h(r)$
representing the solutions of the $M=T=1$ sector approach the profile
of the monopole solution (i.~e. $M=1,T=0$). At the same time, the
function $f(r)$ is more and more peaked at $r=0$ (in particular
$\lim_{\kappa \rightarrow 0} f'(0)=\infty)$ and tends to zero if $r\not= 0$.

This result demonstrates in particular that the coupling of the Skyrmion to
a monopole cannot stabilise the Skyrmion; the Skyrme term is necessary to
guarantee a localized structure to the $T=1$ soliton.

The same phenomenon occurs with the branch of the gauged Skyrmion
($M=0,T=1$) \cite{BT}. The energy in this limit tends to zero, namely to
the energy of the vacuum ($M=T=0$).

A remark should be made concerning the interpretation of the 
monopole-Skyrmion as a bound system of a monopole with magnetic charge 
$M$ and a gauged Skyrmion with baryon number $T$.
Consider a monopole located in a region $U_{\rm m}$
centered at a point $x_{\rm m}$ 
and a gauged Skyrmion located at in a region $U_{\rm Sk}$ centered 
at a point 
$x_{\rm Sk}$ far away from each other.
Then the Skyrmion field and the corresponding gauge field will 
vanish outside the region $U_{\rm Sk}$. 
Consequently, $U_{\rm m}$ contains a pure monopole, consisting of
a gauge field and a Higgs field.
Outside $U_{\rm m}$ the gauge field will vanish, however,
the Higgs field does not vanish. 
Instead it will be equal to its VEV $\langle \Phi \rangle_{\rm vac}$.
In the region $U_{\rm Sk}$ containing the Skyrmion the non-zero
Higgs field is still present and 
we have to allow for interaction with the gauge field,
$|D_j \langle \Phi \rangle_{\rm vac}^\alpha |^2$.
The Higgs vacuum is a constant far away from the monopole and 
generates masses for the gauge fields. Furthermore, the 
Higgs vacuum breaks the rotational symmetry. Consequently, the 
gauged Skyrmion solutions in the presence of a constant Higgs field
will no longer possess spherical symmetry. 
In addition, it might be expected that the electro-magnetic flux 
will not vanish.
If we impose  the condition for the Higgs vacuum that the interaction with
the gauge field has to vanish, 
$D_j \langle \Phi \rangle_{\rm vac}^\alpha = 
\epsilon^{\alpha\beta\gamma}A_j^\beta 
\langle \Phi \rangle_{\rm vac}^\gamma = 0$,
then we will find that the gauge field has to be parallel to the 
Higgs vacuum in isospace. This also breaks the spherical symmetry.

To conclude, the interpretation of the monopole-Skyrmions
as a bound state of a spherically symmetric monopole and 
a spherically symmetric gauged Skyrmion seems to 
be misleading.

\section{Numerical results, case ${\bf A}_0 \neq 0$}

In order to obtain a non trivial function $g(x)$ from eq.~(\ref{eqc}), a non
vanishing asymptotic value, say $q$, for this function has to be imposed
\cite{jz}.
In the asymptotic region $x\gg 1$ eq.~\re{eqc} is satisfied by \be 
\label{dyas}
g(x)
= q - {c_1 \over x} + {\rm o}(x^{-2})
\ee
where $q, c_1$ are constants and $q$ plays a major role in the
construction. The equations, together with the finite energy condition
require $ 0 \leq q \leq 1$,  which can be seen as follows.
In eq.~(\ref{eqa})
the Higgs field and the dyon field contributions, $h^2(x)-g^2(x)$,
generate asymptotically the mass term $m^2_{(a)}=1-q^2$ for the gauge field
function $a(x)$. For $q>1$, $m^2_{(a)}$ becomes negative and leads to an
oscillating function $a(x)$ in the asymptotic region.
Consequently, the term $a^2 g^2$ in eq.~(\ref{ec}) is not integrable and 
no dyon solution exists for $q>1$.

 The electric charge, as defined
in \cite{jz}, is directly related to the constant $c_1$~:
\begin{eqnarray}
Q &=& \frac{1}{4\pi \eta} \int \vec \Phi \cdot\vec F_{0i}
\: dS_i \equiv {1 \over e} \tilde Q \\
&=& \frac{1}{4\pi} \int (r^2 \frac{dg}{dr})|_{r\to \infty}
\sin \theta d\theta d\phi ={1 \over e} c_1 ,\label{dyas1}
\end{eqnarray}
having used the Ansatz \re{3.1}-\re{3.2} and eq.~\re{dyas}.

Another very interesting quantity is the chiral anomaly due to the
dyon-Skyrmion soliton whose classical solutions will be studied numerically.
The anomaly equation for the chiral charge is
\begin{eqnarray}
\frac{dQ_5}{dt} &=& \frac{e^2}{8\pi^2} \int d {\bf x}
\: {\bf E}_i \cdot{\bf B}_i \nonumber \\
&=& -\frac{e^2}{8\pi^2} \: 4\pi \:
[g(r)(a(r)-1)]_{r=0}^{\infty} = \frac{e^2}{2\pi} q \label{cp},
\end{eqnarray}
having used the Ansatz \re{3.1}-\re{3.2} and, eq.~\re{dyas}. 
We now discuss the
solutions by adopting the same presentation as in the previous Section.

\subsection{Case $M=1,T=0$}
The solutions are the dyons of Julia and Zee \cite{jz}. Here we present an
in depth analysis of this solution. The limit $\lambda = 0$ corresponds to
the Prasad-Sommerfield dyon \cite{ps} (PS dyon). It is worth analysing this
case separately because the solution can be computed analytically
and it provides a good check of our numerical routines.

\vspace{0.5cm}
\noindent {\bf Case $\lambda = 0$}
\vspace{0.5cm}

The profile of the radial functions of the PS dyon reads \cite{ps}
\begin{eqnarray}
& a(x) &= {cx \over \sinh(cx)} \\
& g(x) &= {cq \over \sqrt{1-q^2}} (\coth cx - {1\over cx})
\\
& h(x) &= {c \over \sqrt{1-q^2}} (\coth cx - {1\over cx})
\end{eqnarray} and the PS monopole is recovered for $q=0$.
Our parameter $q$ is related to $\gamma$ of Ref.~\cite{ps}, by
$q = \tanh(\gamma)$. We have chosen the arbitrary scale in the 
PS solution
$c=\sqrt{1-q^2}$ so that the asymptotic value of the Higgs 
field function $h(x)$
of the PS solution given above be
equal to $1$, since we are also studying the dyons for the $\lambda >0$ 
case
where the asymptotic value of $h(x)$ equals $1$.
The charge and energy of the PS dyon are given by
\be
\tilde Q = {q \over \sqrt{1-q^2}} \ \ , \ \ \tilde E = { 1 \over
\sqrt{1-q^2}} \label{dyonch}
\ee
\be
\tilde E_1 = { q^2 \over 2\sqrt{1-q^2}} \ \ , \ \ \tilde E_2 = (1-{1\over
2}q^2){ 1 \over \sqrt{1-q^2}} \approx 1 + {1\over 8} q^4 + o(q^{-6})
\label{dyonen}
\ee
For small values of $q$
the "magnetic" contribution to the energy, $E_2$, varies slightly with
$q$, accounting for the feed back of the electric charge on the classical
magnetic energy. We would like to stress that our numerical results are in
full agreement with these exact formulas.

The dependence of the charge $\tilde Q$ of the PS dyon as a function of $q$
is represented in Fig.~6 (curve $a$). Similarly we have reported on Fig.~7
(curve $a$) the energy $\tilde E$ of the PS dyon as a function of $\tilde
Q$. Clearly the energy and the charge of the PS dyon can be arbitrarily
large when $q \rightarrow 1$.

\vspace{0.5cm}
\noindent {\bf Case $\lambda \neq 0$}
\vspace{0.5cm}

For the dyon solution, the boundary conditions for the function $g(x)$ 
can
be read from eq.~(\ref{dyas}), and those of the functions $a(x)$, $h(x)$ 
from
eqs.~(\ref{moor}) and (\ref{moas}), with the exception of the behaviour 
of the
function $a(x)$ in the $x\gg 1$ region, which now takes the form \be
\label{dyaas}
a(x) \simeq A e^{-\sqrt{1-q^2}x} \ .
\ee
The main distinguishing feature of the $\lambda \neq 0$ dyon vs. the PS
dyon is that its electric charge and its classical energy are bounded for
$q \in [0,1]$.

This phenomenon appears clearly in Fig.~6 and Fig.~7 respectively, 
where the
quantities
$\tilde Q$ as a function of $q$, and $\tilde E$ as a function of 
$\tilde Q$, are plotted for $\lambda = 0.5$.
More generally, it appears that the
electric charge of the dyon constructed with a given value of the parameter
$q$ decreases when $\lambda$ increases. The three bullets on Fig.~7
represent the data given in \cite{jz}; according to our numerical results
they should lie on line $b$.
Our numerical results therefore slightly desagree with \cite{jz}.

The star on line $b$ of Fig.~7 indicates the maximal accessible charge of
the dyon solutions for a fixed value of $\lambda$. This contrasts with line
$a$ which asymptotically tends to infinity, in agreement with
eqs.~(\ref{dyonch}) and (\ref{dyonen}). The solutions with maximal 
electric charge
and energy correspond to the case $q=1$ which we discuss next.

\vspace{0.5cm}
\noindent {\bf Case $q = 1$}
\vspace{0.5cm}

In the limit $q=1$ in eq.~(\ref{dyas}) the equation (\ref{eqa}) ceases to
impose the exponential decay
eq.~(\ref{dyaas}) for the function $a(x)$; we have instead 
\be a(x) \simeq A
e^{-\sqrt{8 c_1 x}} \ \ \ {\rm for} \ \ x\rightarrow \infty \ee 
where $c_1$
is defined in eq.~(\ref{dyas}).

Fixing $\lambda \neq 0$, the
electric charge (and similarly the classical energy) of the dyon cannot
exceed a critical value, say $Q_{cr}(\lambda)$. This quantity is plotted
against $\lambda$ on Fig.~8 (solid line).

\subsection{Case $M=0,T=1$}
No finite energy dyon-like solutions supporting a non-vanishing
(non-Abelian) electric field can be found in this case. Due to the
 absence
of the Higgs field ($h=0$), eq.~(\ref{eqa}) leads to an oscillating
asymptotic behaviour of $a(x)$. The term $a^2 g^2$ in the energy 
eq.~(\ref{ec})
can therefore not be integrated.

\subsection{Case $M=1,T=1$}
The boundary conditions compatible with a finite energy solution in this case
are identical to eqs.~(\ref{skmoas1}), (\ref{skmoas2}) and (\ref{dyas}). 
It is possible to construct the dyon-Skyrmion solutions. 
The dyon-Skyrmion display many
features of the dyons, discussed at length above. These features are 
illustrated
by Figs.~6 and 7 (dashed curves $c$, $d$ and $e$) and by Fig.~8 (dashed line). 
In addition we illustrate the dependence of the energy on the Skyrme coupling 
constant $\kappa$ in Fig.~9 (solid line) for $q=0.5$ and $\lambda = 0$.
The energy is an increasing function of $\kappa$. In the limit of vanishing 
$\kappa$ the energy of the dyon-Skyrmion converges to the energy of the 
dyon-monopole. This can be compared with the behaviour of the energy of the 
monopole-Skyrmion, shown (for $\lambda = 1$) in Fig.~1, where for vanishing 
$\kappa$ the energy tends to the energy of the monopole.
%In
%addition,
%we illustrate the dependence of the energy of the dyon-Skyrmion in
%the case with $q=0.5$, on the Skyrme coupling constant $\kappa$ in Fig.~10
%(solid curve). The latter is the analogue Fig.~1.~2,
%where the dependence of the energy of the monopole-Skyrmion was plotted 
against
%$\kappa$, and like the latter the energy of the dyon-Skyrmion sinks to the
%value of the dyon when $\kappa =0$. (The only qualitative difference between
%Fis. 2 and 10 is that in Fig.~1 we have set $\lambda =1$ while in Fig.~10
%$\lambda =0$.) While in Fig.~1 we displayed the energy
%curve for the gauged-Skyrmion on its own however, here in Fig.~10 the
%analogue of that curve is absent as there do not exist any
%finite enrgy solutions of that system which in the ${\bf A_0}=0$ gauge 
support a
%nonvanishing (non-Abelian) electric field. 
For our considerations leading to our conclusions in Fig.~9, 
we have chosen $q=0.5$ as a typical value in
the allowed  range $0\le q\le 1$. 
We expect that our results, summarised by the
solid curve in Fig.~9, is typical for any allowed value of $q$, and also 
for any
value of the Higgs coupling constant $\lambda$, except in the important case
of $\lambda =0$ and $q=1$. The dyon-Skyrmion characterised by the boundary
value $q=1$
in the $\lambda =0$ model has peculiar and interesting properties which we
analyse in the next paragraphs.

For $\lambda = 0$ the solutions of eqs.~(\ref{eqc}) and (\ref{eqh}) 
are proportional to each other. 
Assuming that $h=1$ at infinity, the proportionality
constant is given by $q$, eq.~(\ref{dyas}). 
Thus for $q=1$ the functions $h(r)$ and $g(r)$ are identical. 
In this special case $h^2(r)$ and $g^2(r)$
cancel each other in eq.~(\ref{eqa}).
Consequently, eqs.~(\ref{eqa}) and
(\ref{eqf}) reduce to the field equation of the gauged Skyrme model (Sec.
4.3), and can be solved independently of $h(r)$, $g(r)$.

The solutions of these equations are now given by the branch B solutions of
the gauged Skyrme model. Once a solution for the function $a(r)$ is found
the equations for the functions $h(r)$ and $g(r)$ can be solved.
Recalling that the branch B solutions exist for all 
$\kappa \geq \kappa_B^{cr}$,
we expect the existence of the dyon-Skyrmion solution for the same range of
coupling constants $\kappa$. However, not all of these solutions are finite
energy solutions. This can be seen easily by inspecting the static
Hamiltonian
${\cal E}={\cal E}_1 +{\cal E}_2$ given in eqs.~(\ref{ed1}) and
(\ref{ed2}), where the contributions from the functions $h(r)$ and $g(r)$
do not cancel. The asymptotic behaviour of these terms is dominated by
$a^2(r) h^2(r)$, $a^2(r) g^2(r)$.
Using the boundary conditions $h(\infty )=1$, $g(\infty )=1$ and the
asymptotic form of the function $a(r)$, eq.~(\ref{gsasb}), these terms
behave like
$ \approx {A^2(\kappa) / x^{2 \alpha(\kappa )}}$ for large $x$, 
where $\alpha(\kappa)$ is a function determined numerically. 
Thus the
integration of these terms will give finite contributions only if
$\alpha(\kappa ) > 1/2$. This restricts the range of the coupling constant
$\kappa$ to $\kappa^{cr}_{1/2} < \kappa < \infty $, where $
\kappa^{cr}_{1/2}$ is defined by $\alpha(\kappa^{cr}_{1/2})=1/2$. For
$\xi=1$ we find $\kappa^{cr}_{1/2}=0.7652$.

In Fig.~9 we show the dependence of the energy on the coupling constant
$\kappa$ for $q=1$ (dotted line) and for $q=0.5$ (solid line).
(For $q<1$, as stated above, solutions exist for all values of $\kappa$, 
the energy is a
monotonic function of $\kappa$, and the limit $\kappa \rightarrow 0$ the
energy approaches the energy of the dyon solution.)
For $q=1$ the energy is an increasing function of $\kappa$ for large values
of $\kappa$ only. It has a minimum at $\kappa = 1.21$. As $\kappa$
approaches its critical value $\kappa^{cr}_{1/2}$ the energy becomes
increasingly large and diverges at $\kappa = \kappa^{cr}_{1/2}$.

The charge $\tilde{Q}$ of the solutions is determinded by the asymptotic
behavior of the function $g(r)$, eq.~(\ref{dyas1}). Solving the equation
(\ref{eqc}) for large $x$ we find the following expressions for the charge
\begin{equation}
\tilde{Q} = \left\{
\begin{array}{lcl}
{\displaystyle \lim_{x\rightarrow \infty} \left( c_1-\frac{2 A^2}{2 \alpha
-1} x^{-(2 \alpha -1)} \right) }
& {\rm for } & {\displaystyle \alpha > \frac{1}{2} }\ , \\ {\displaystyle
\lim_{x\rightarrow \infty} \left( 2A^2 \ln(x) \right) }
& {\rm for } & {\displaystyle \alpha = \frac{1}{2}}\ , \\ {\displaystyle
\lim_{x\rightarrow \infty} \left( -\frac{2 A^2}{2 \alpha -1} x^{-(2 \alpha
-1) }\right)}
& {\rm for } & {\displaystyle \alpha < \frac{1}{2}}\ . \end{array}
\right.
\end{equation}
Thus solutions with finite charge exist only for $\alpha > 1/2$, i.~e. for
the same range of the coupling constant $\kappa$ where finite energy
solutions exist.

In Fig.~6 we show the dependence of the charge on the parameter $q$ for
$\kappa = 0.4$ and $\kappa = 1.0$.
For $\kappa = 0.4 \ (< \kappa^{cr}_{1/2}) $ there is no finite charge 
solution
for $q=1$. Consequently,
the charge as a function of $q$ diverges as $q$ approaches the value $1$.
In contrast, for $\kappa = 1 \ (> \kappa^{cr}_{1/2} )$ the solution 
with $q=1$
exists and the charge is finite for all values of $ q \in [0,1]$.

In Fig.~7 the energy as a function of the charge is shown for 
$\kappa = 0.4$
and $\kappa = 1.0$.
For $\kappa = 0.4 \ (< \kappa^{cr}_{1/2}) $ the energy and the charge 
can take
arbitrarily large values. In this case the energy is a monotonically
increasing function of the charge with no end point. 
For $\kappa = 1 \ (> \kappa^{cr}_{1/2}) $ 
the energy is again a monotonically increasing function
of the charge.
However, only finite energy and charge solutions exist for this value of
$\kappa$.
Thus the graph of the function $\tilde{E}(\tilde{Q})$ ends 
at the maximal value of the charge.

\section{Summary and Discussion}

We have found monopole-Skyrmion and dyon-Skyrmion solutions to an $SO(3)$
gauged Higgs and $O(4)$ sigma (Skyrme) model, in which both scalar matter
fields interact with the gauge field but not with each other. The Higgs
field is isovector,
like in the GG model, while the $S^3$ valued (sigma) field is gauged
according to the prescription used in Refs. \cite{AT,BT}.

In the $A_0^{\alpha} =0$ gauge the static Hamiltonian is bounded from below
by the sum of the two topological charge densities, the monopole charge and
the degree of the map of the $S^3$ field on $\bfR_3$.Thus the imposition of
spherical
symmetry reduces the system to an one dimensional subsystem, and the
resulting differential equations are integrated analytically in the
asymptotic regions and then numerically. This yielded the
monopole-Skyrmion.

In the $A_0^{\alpha} \not = 0$ gauge, the Euler-Lagrange equations arising
from the variation of the static Hamiltonian density do not yield a soliton
with non-vanishing $A_0^{\alpha}$ and hence have $E_0^{\alpha} =0$. 
Instead, the variational equations arising from the 
(non positive-definite) Lagrangian density in the static limit
support spherically symmetric solutions with $E_i^\alpha \neq 0$. 
This is also what happens with the JZ dyon.
There \cite{jz},
inspite of the non--positive-definiteness of the functional subjected to
the variational principle, it happens that after taking the static limit
and imposing spherical symmetry, these equations reduce to a set of
consistent, i.~e. not overdetermined,
set of coupled ordinary differential equations. Their solutions support a
non-vanishing $A_0^{\alpha}$ field. These ordinary differential equations
also result
from the variation of a certain one-dimensional (radial) functional which,
in contrast to the one dimensional energy functional, is not positiv 
definite.

In the light of the surprisingly successful outcome for the JZ dyon, we were
motivated to address the same question for the $SO(3)$ gauged $O(4)$ model
\cite{AT,BT}. Subjecting the Lagrangian to the variational principle and then
taking the static limit and imposing spherical symmetry, we found that this
also led to a consistent set of coupled ordinary differential equations.
The same situation obtains with the composite model of this paper, 
and it is the
dyon like
soltions of these last equations which yielded the dyon-Skyrmion.
Concerning the $SO(3)$ gauged Skyrme model
on its own, while its equations of motion reduce to a consistent set of
coupled ordinary differential equations, their solutions support only
vanishing electric field.

As a byproduct of our study of the dyon-Skyrmion, we made a detailed
re-analysis of the JZ dyon refining our understanding of the latter, for
example exploring the
dependence of the energy of the dyon on its electric charge. 

An important result of the numerical analysis of the monopole-Skyrmion
 solution
is that, as the coupling strength of the Skyrme term is shrunk down to zero
the monopole-Skyrmion reduces to the monopole, as depicted in Fig.~1.
Thus the monopole does not stabilise the $SO(3)$ gauged sigma model
 without a
Skyrme term, something that is not prohibited by the Derrick scaling
requirement.

Perhaps the most interesting aspect of the dyon-Skyrmion occurs for the
model in the PS limit ($\lambda =0$) in the special case where the
boundary value
$q$ of the function $g(r)$ parametrising $A_0^{\alpha}$ equals $1$. 
In this case,
the equations governing the functions $a(r)$ (parametrising the gauge
field) and the
function $f(r)$ (governing the Skyrme field)
decouple from the fields $h(r)$ (parametrising the Higgs field) and $g(r)$. 
As a consequence the solutions for the functions $a(r)$ and $f(r)$
are just the (branch B of the) gauged-Skyrmion solutions
and exist only for
values of the Skyrme coupling constant larger than a critical value 
$\kappa^{cr}_B$, as seen from Fig.~2. 
However, when the integrations of the Higgs field function $h(r)$
and of the dyon function $g(r)$ are taken into account,
then finite energy solution only exist
if the Skyrme coupling constant is larger than the critical value
$\kappa^{cr}_{1/2}>\kappa^{cr}_B$, see Fig.~9.
The energy of the solution at this critical value is found to become 
infinite
and for lower values of the Skyrme coupling constant no
finite energy solution exists.
The time rate of change of the chiral charge eq.~\re{cp} is
equal to the integer $1$ (up to normalisation) for all values of the 
Skyrme
coupling constant
$\kappa$ down to the critical value $\kappa_{1/2}^{cr}$, below which 
no finite energy solutions exist.
We hope that this result may prove relevant to the semiclassical description
of monopole catalysis of Fermion number non-conservation.  
If for example it could be argued that the dyon-Skyrmion favoured by Nature 
is the solution to the system eq.~(\ref{2.5}) in the PS limit, with the 
asymptotic constant $q=1$, i.~e. for which the quantity 
${\displaystyle \frac{d Q_5}{dt}}$ is an integer (up to normalisation),
then it would follow that below the critical value $\kappa^{cr}_{1/2}$ 
there will be no $Q_5$ violating rate.
We intend to return
to this question in the near future.

\bigskip

{\bf \large Acknowledgements}
We are greatful to V.~A. Rubakov for helpful discussions. B.~K. was
 supported 
by Forbairt grant SC/97-636, and Y.B. and D.~H.~T. acknowledge partial 
support from Forbairt project IC/97/019.

%%%\newpage
\small{

 }

\newpage
\begin{table}[p!]
\begin{center}
\begin{tabular}{|c|ccc|} 
 \hline 
          & monopole & monopole-Skyrmion     & gauged Skyrmion \\
$\lambda$ &          & ($\xi=1$, $\kappa=0.4$) & ($\xi=1$, $\kappa=0.4$) \\
 \hline 
0.0  &	1.000 &	3.450 &	2.98 \\
0.05 &	1.106 &	3.470 &	2.98 \\
0.10 &	1.138 &	3.480 &	2.98 \\
0.20 &	1.180 &	3.490 &	2.98 \\
0.40 &	1.220 &	3.510 &	2.98 \\
0.60 &	1.250 &	3.520 &	2.98 \\
0.80 &	1.270 &	3.530 &	2.98 \\
1.00 &	1.290 &	3.536 &	2.98 \\
 \hline 
\end{tabular}
\end{center} 
\vspace{1.cm} 
{\bf Table 1}\\
The energies of the monopole, the monopole-Skyrmion and the gauged Skyrmion
for several values of the Higgs coupling constant $\lambda$.
\end{table}
\clearpage
\newpage
\centerline{Figure Captions}
\begin{itemize}

\item [Figure 1] The energies eq.~(\ref{emon}) of the monopole (line $a$), 
of
the gauged Skyrmion (line $b$) and of the monopole-Skyrmion (line $c$) as
functions of $\kappa$ ($\lambda =1, \xi=1$).

\item [Figure 2]
The energy of the gauged Skyrmion as a function of $\kappa$ in the region
of the phase transition.
The branches $A$, $\tilde A$ are represented by the solid line and branch
$B$ by the dashed line.

\item [Figure 3]
The (logarithm of the) function $a(x)$ on the two branches $B$ and $\tilde
A$ on a logarithmic scale for several values of $\kappa$ approaching the 
critical
value $\kappa_B^{cr}$.

\item [Figure 4]
The quantity $F_1$ defined in eq.~(\ref{gsor}) is plotted as a function of
$\kappa$ for the branches $A, \tilde A$ (solid line) and for the branch $B$
(dashed line).

\item [Figure 5]
The quantities $\ln(F-1)$ and $\ln(\alpha)$ (defined in eq.~(\ref{gsasb})) 
are
plotted as functions of the parameter $\ln(\kappa - \kappa_B^{cr})$.

\item [Figure 6]
The values of the electric charge $\tilde Q$ as a function of the parameter
$q$. The solid lines represent the dyon for $\lambda = 0$ (line $a$) and
$\lambda = 0.5$ (line $b$). The dashed lines represent the
dyon-Skyrmion ($\xi=1$) for 
 $\lambda = 0$, $\kappa=0.4$ (line $c$), $\lambda = 0.5$, $\kappa=0.4$ 
 (line $d$)
and $\lambda = 0$, $\kappa=1$ (line $e$).

\item [Figure 7]
The values of the energy $\tilde E$ as a function of the parameter 
$\tilde Q$. The
solid lines represent the
dyon for $\lambda = 0$ (line $a$) and $\lambda = 0.5$ (line $b$). The
dashed lines represent the dyon-Skyrmion ($\xi=1$)  for
$\lambda = 0$, $\kappa=0.4$  (line $c$),
$\lambda = 0.5$, $\kappa=0.4$ (line $d$) and $\lambda = 0$, $\kappa=1$ 
(line $e$). 
The stars depict the points where the solution has maximal finite charge. 
The bullets correspond to the data of \cite{jz}.

\item [Figure 8] The value of the critical charge $Q_{cr}$ as a function of
$\lambda$. The solid line refers to the dyon solution.
The dashed line refers to the 
dyon-Skyrmion solution for $\xi=1$ and $\kappa=0.4$.

\item [Figure 9] The energies of the dyon-Skyrmions with $q=1$ (solid line)
and $q=0.5$ (dashed line) as functions of $\kappa$ ($\lambda$ =0, $\xi$=1).
The vertical dotted line indicates the critical value of $\kappa$. 

\end{itemize}

%\end{document}
\newpage
\begin{figure}
\centering
\mbox{\epsfysize=12.cm\epsffile{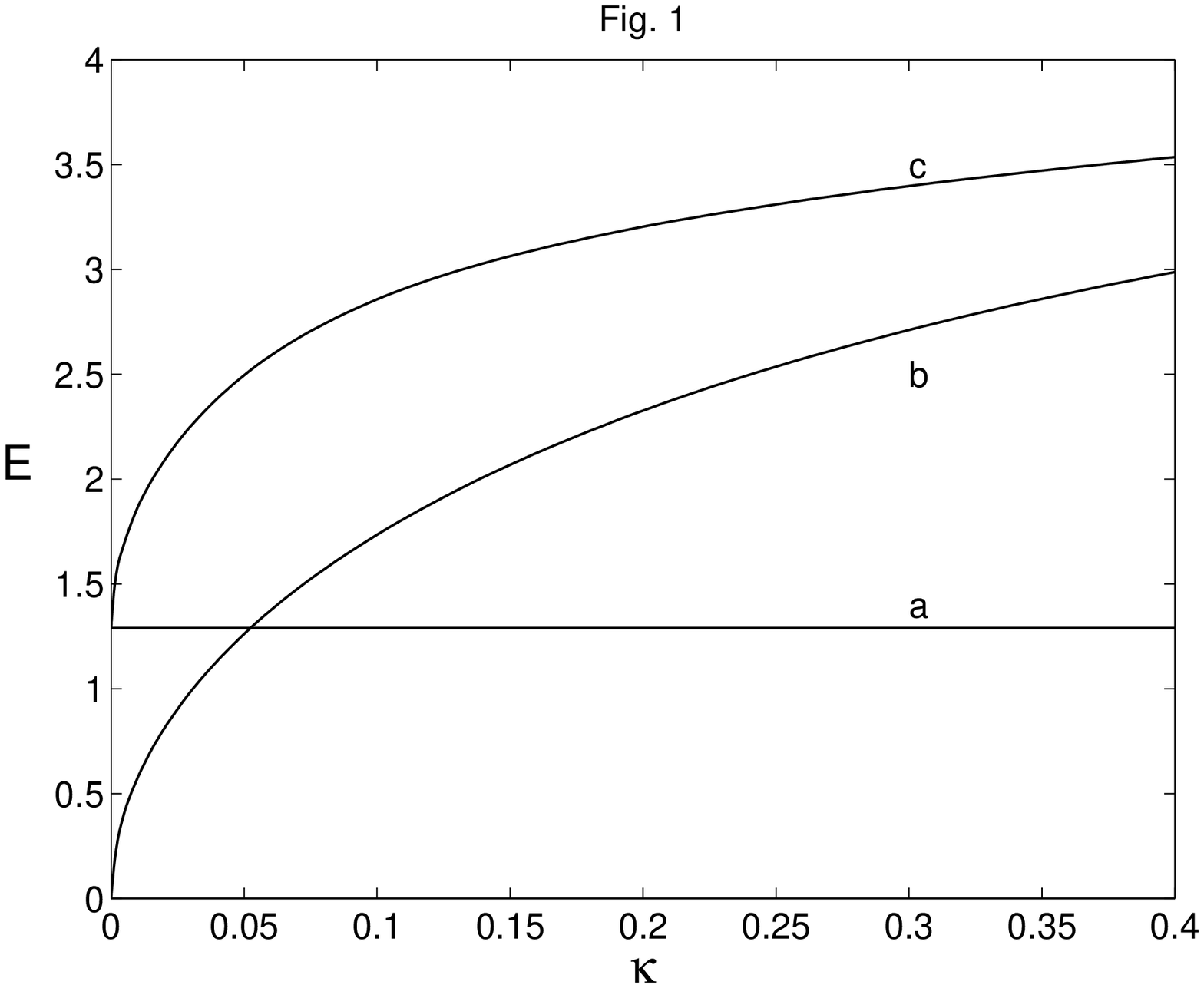}}
\end{figure}

\newpage
\begin{figure}
\centering
\mbox{\epsfysize=12.cm\epsffile{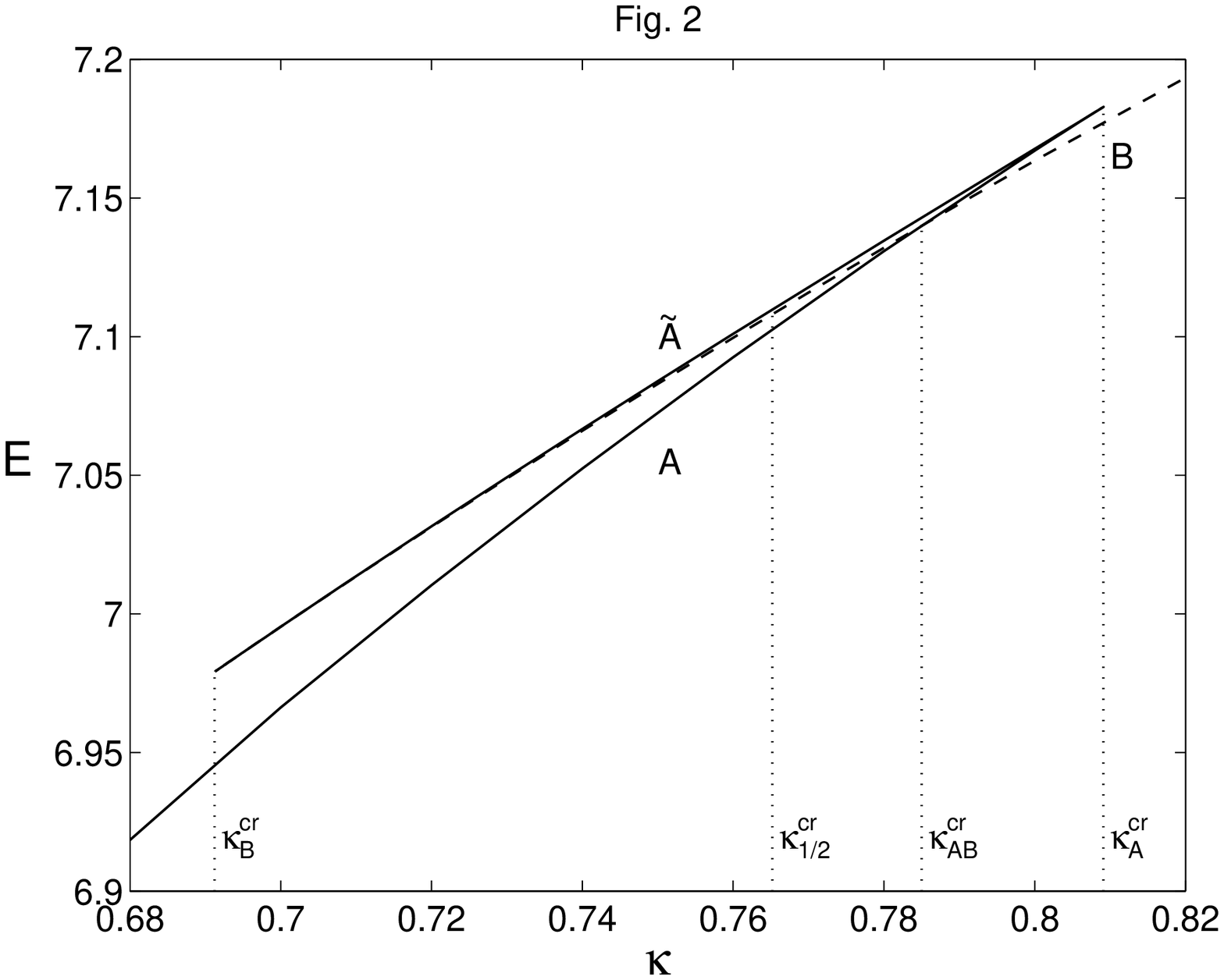}}
\end{figure}

\newpage
\begin{figure}
\centering
\mbox{\epsfysize=12.cm\epsffile{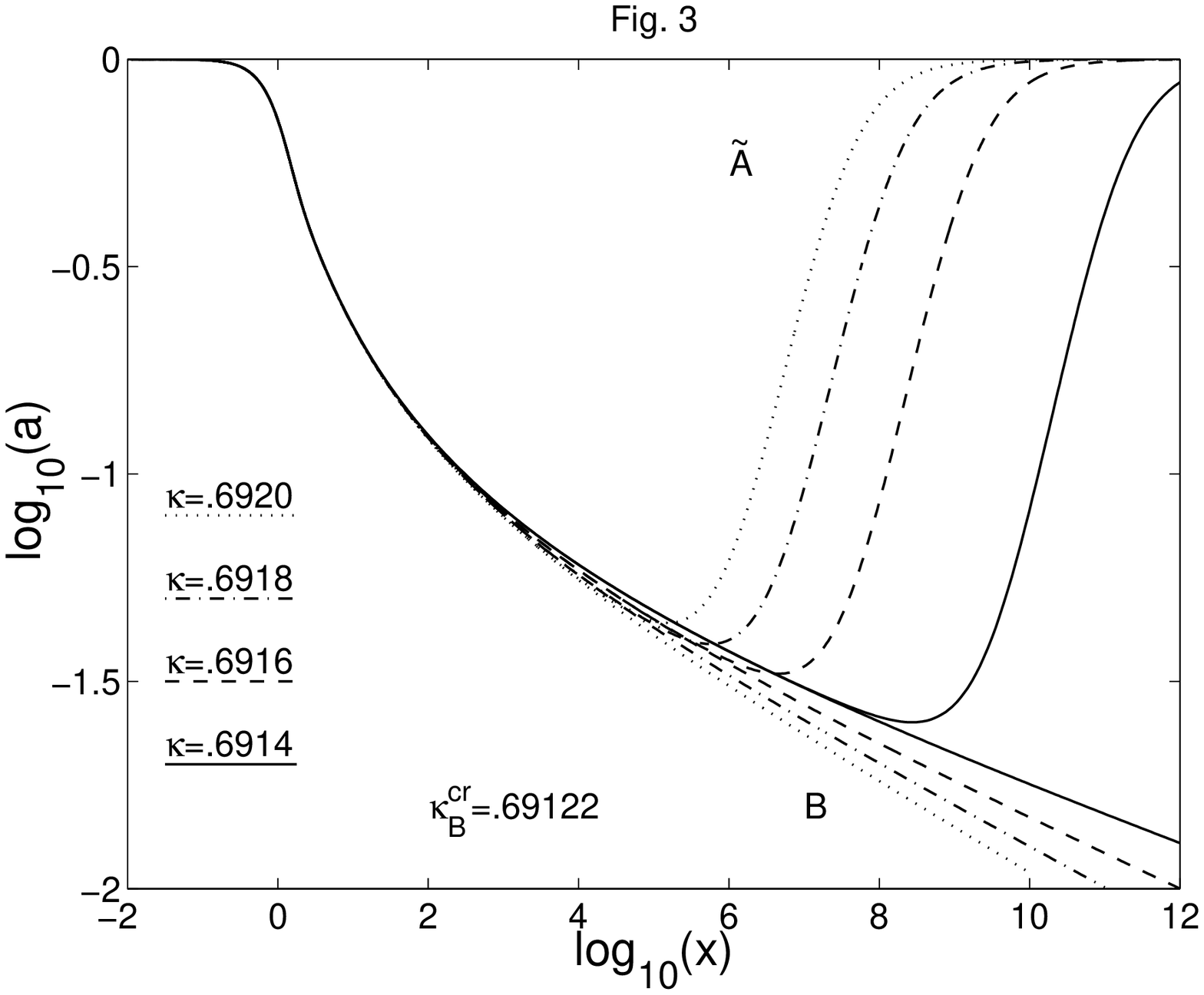}}
\end{figure}

\newpage
\begin{figure}
\centering
\mbox{\epsfysize=12.cm\epsffile{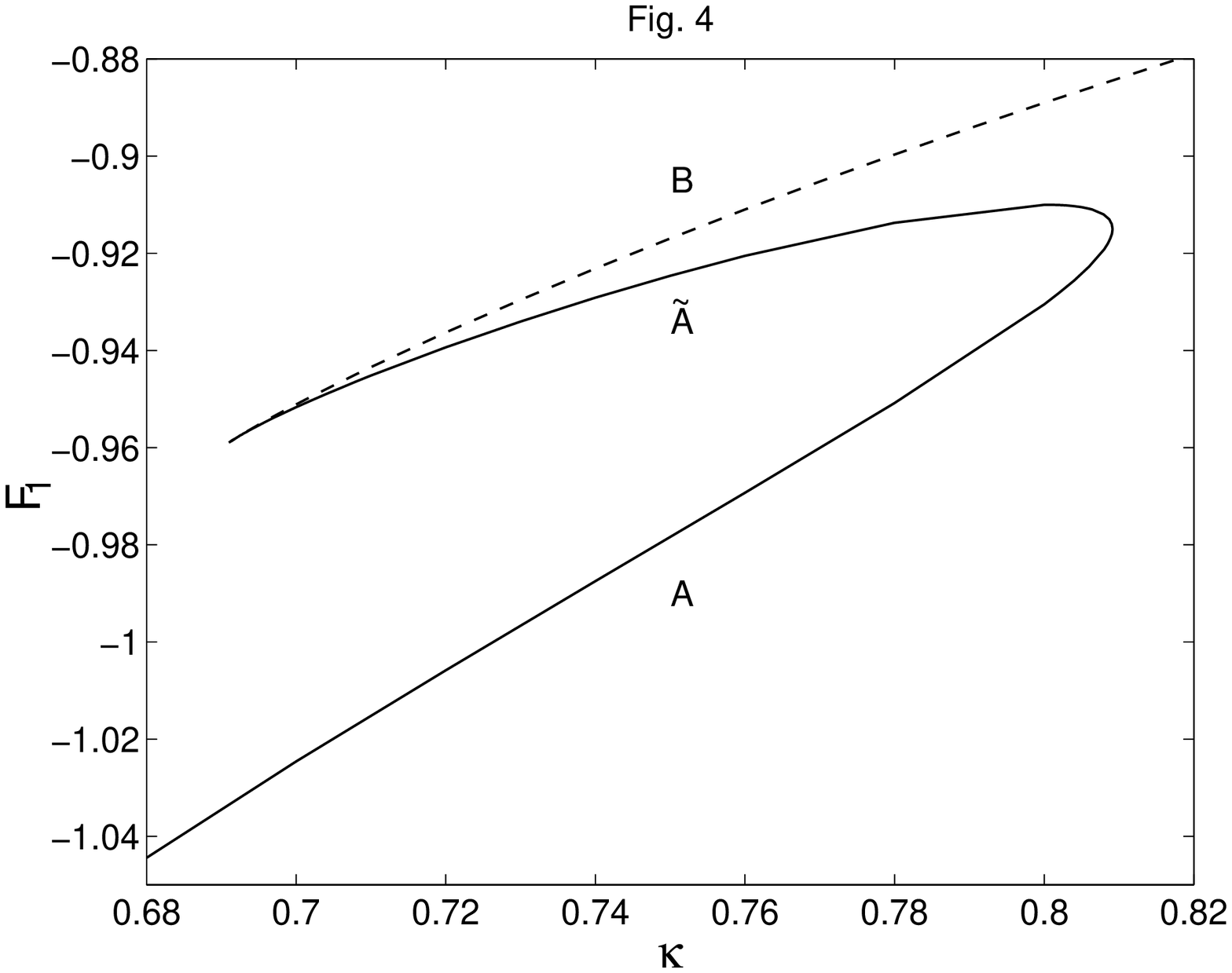}}
\end{figure}

\newpage
\begin{figure}
\centering
\mbox{\epsfysize=12.cm\epsffile{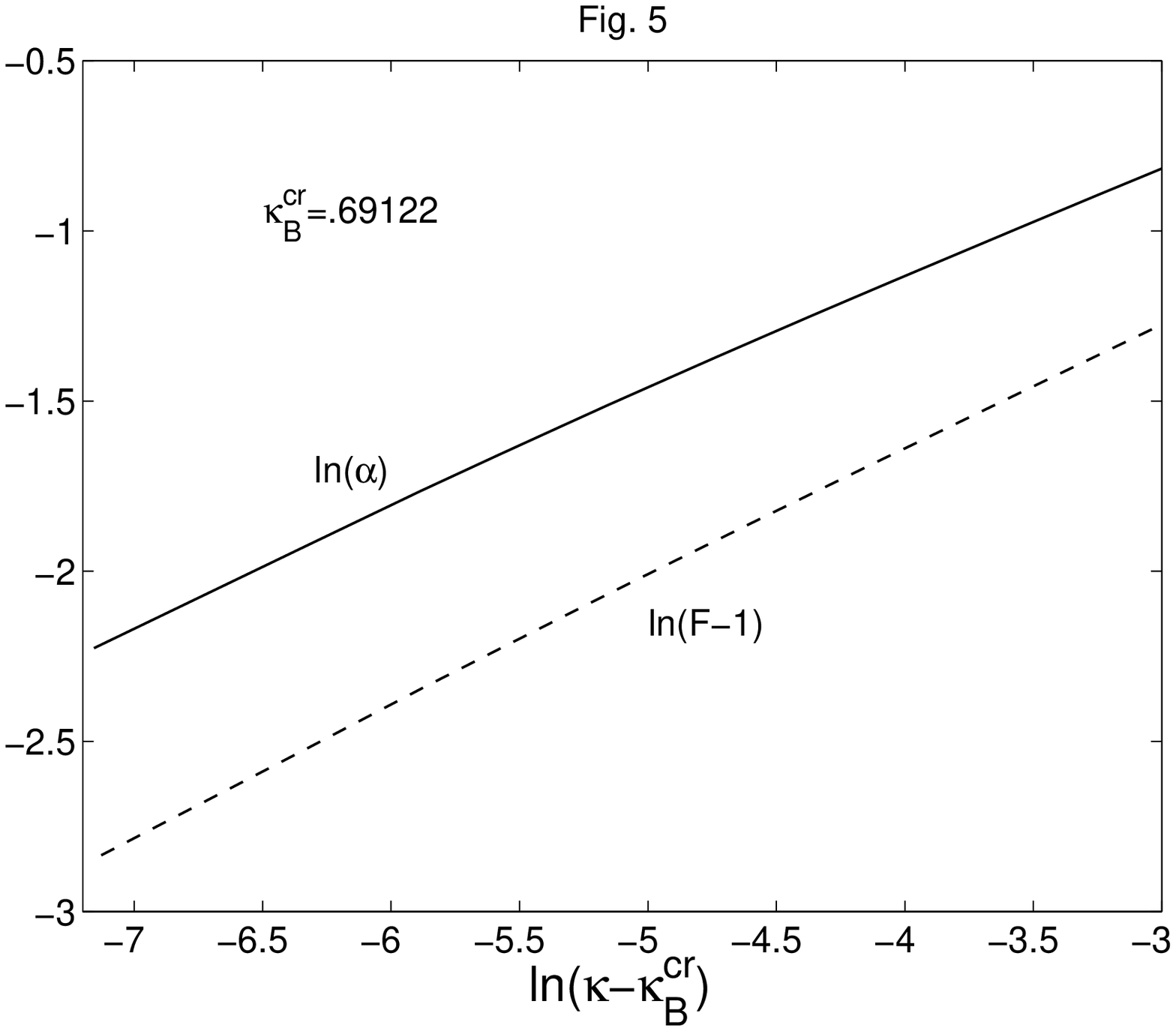}}
\end{figure}

\newpage
\begin{figure}
\centering
\mbox{\epsfysize=12.cm\epsffile{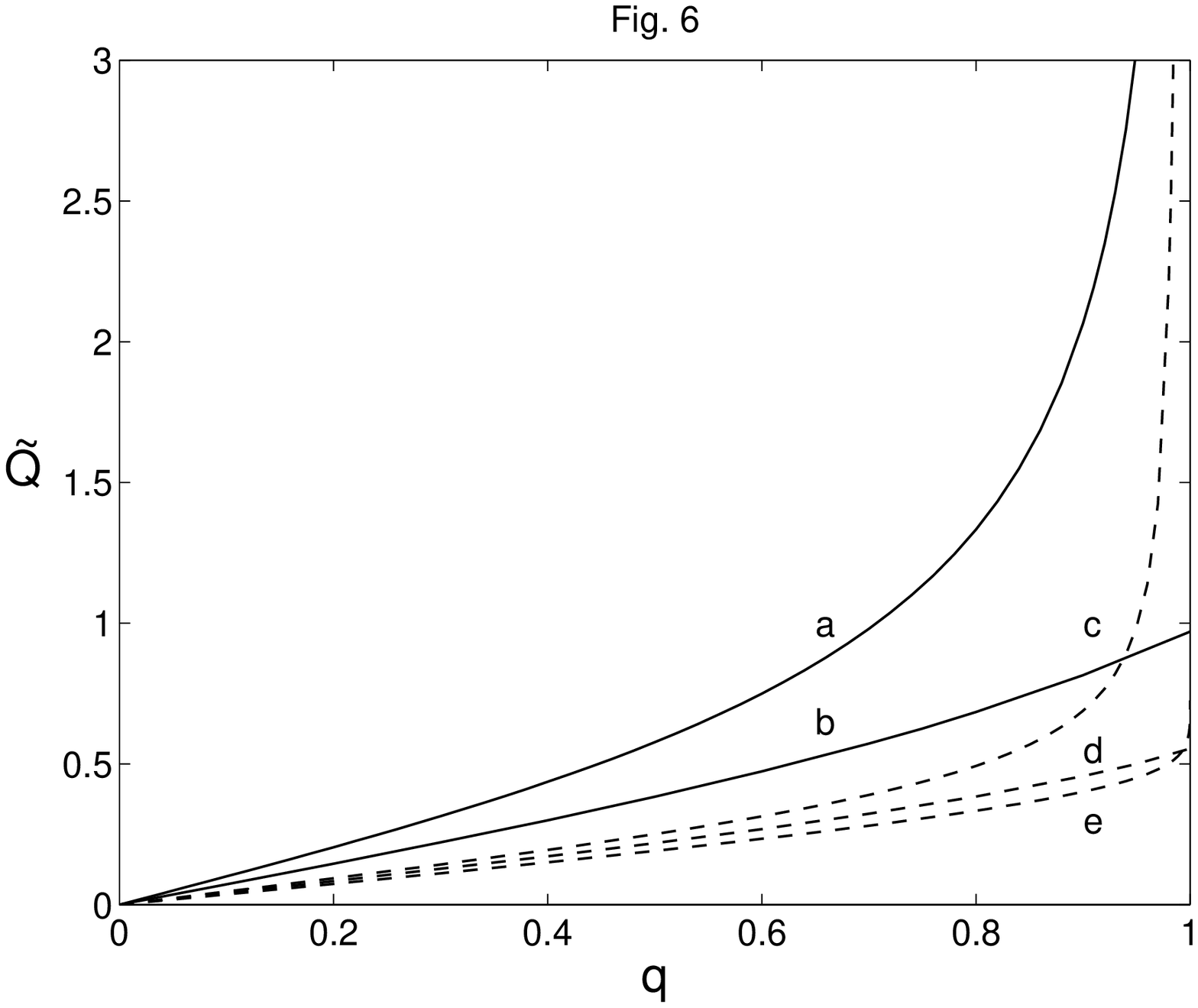}}
\end{figure}

\newpage
\begin{figure}
\centering
\mbox{\epsfysize=12.cm\epsffile{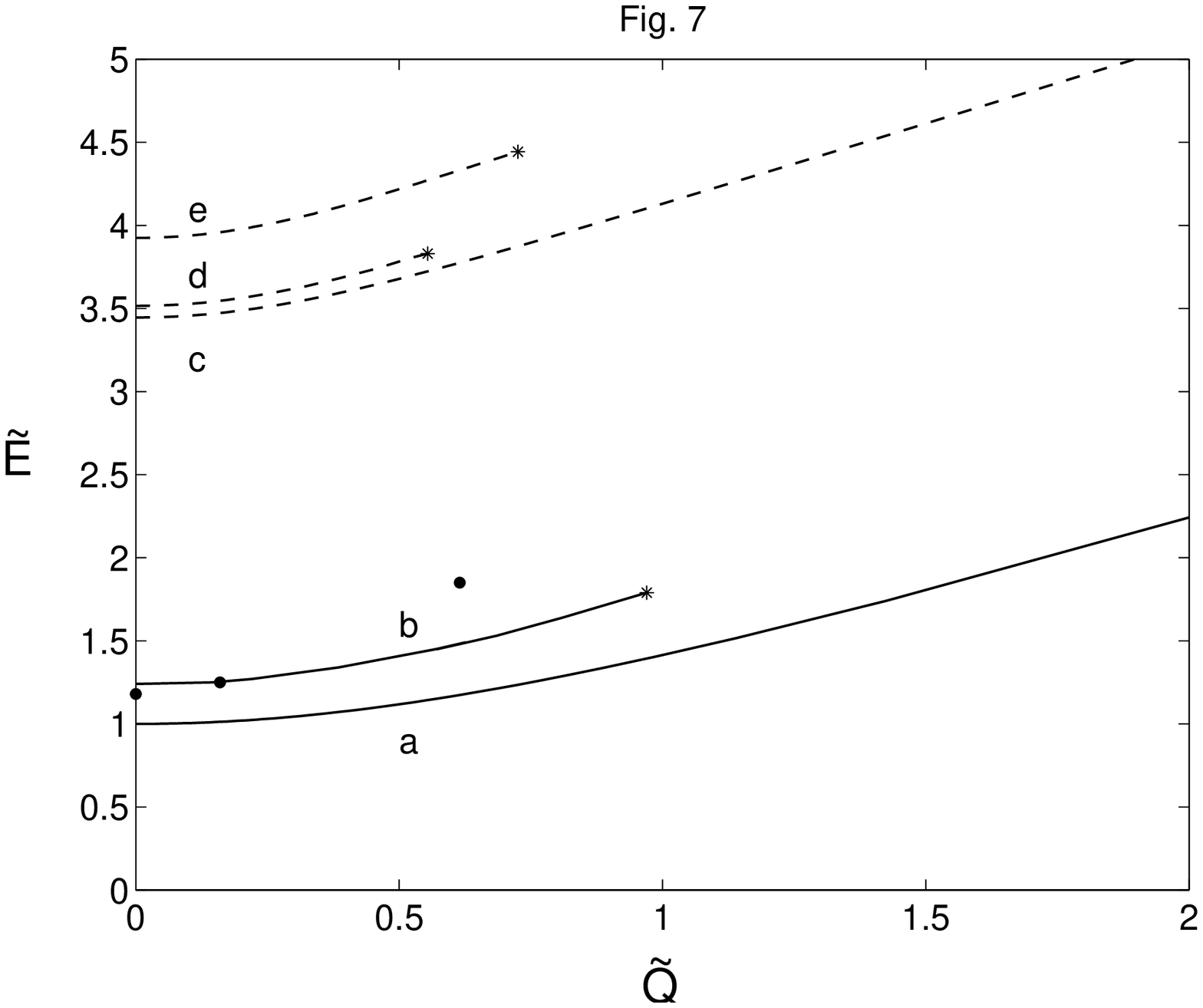}}
\end{figure}

\newpage
\begin{figure}
\centering
\mbox{\epsfysize=12.cm\epsffile{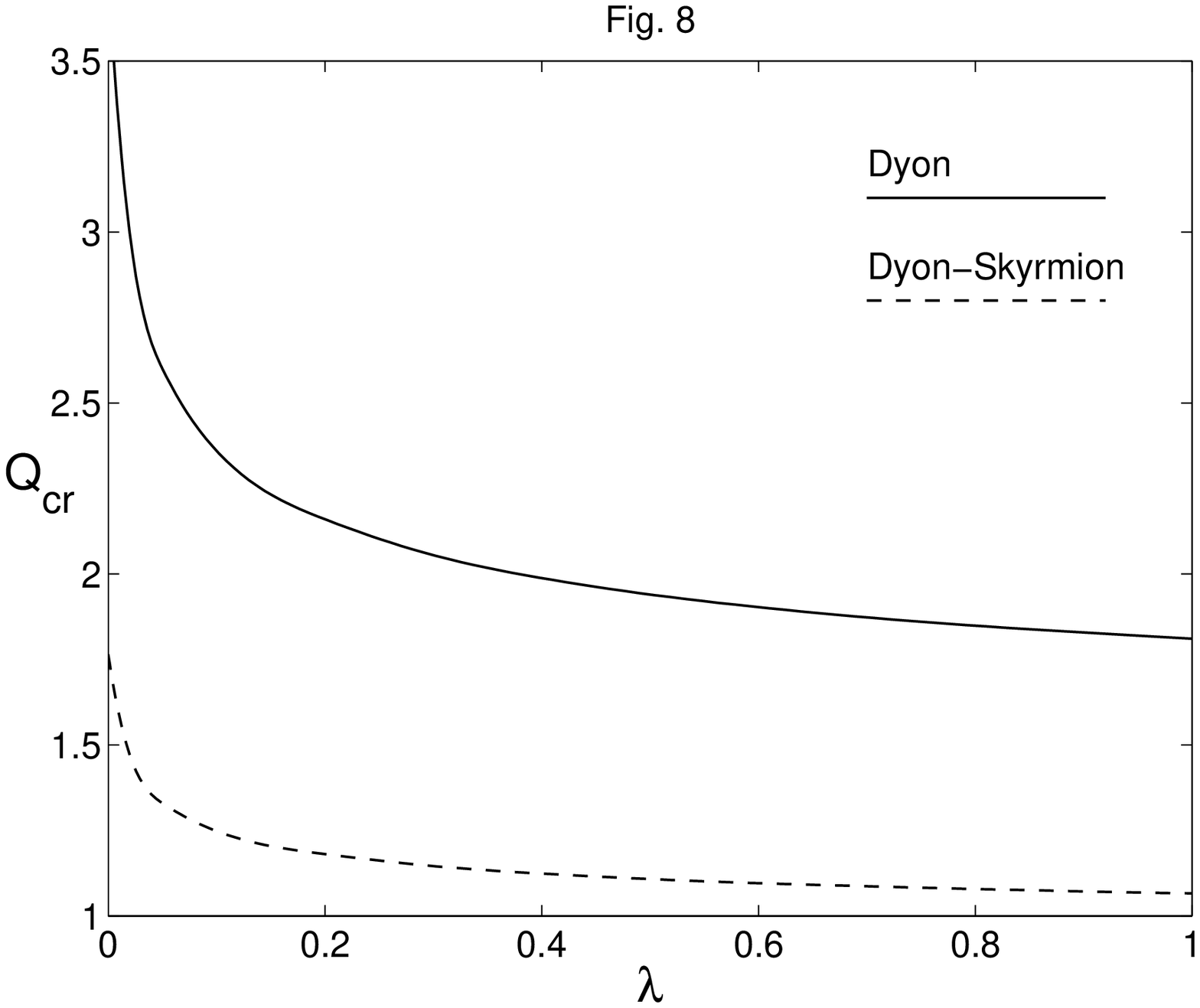}}
\end{figure}

\newpage
\begin{figure}
\centering
\mbox{\epsfysize=12.cm\epsffile{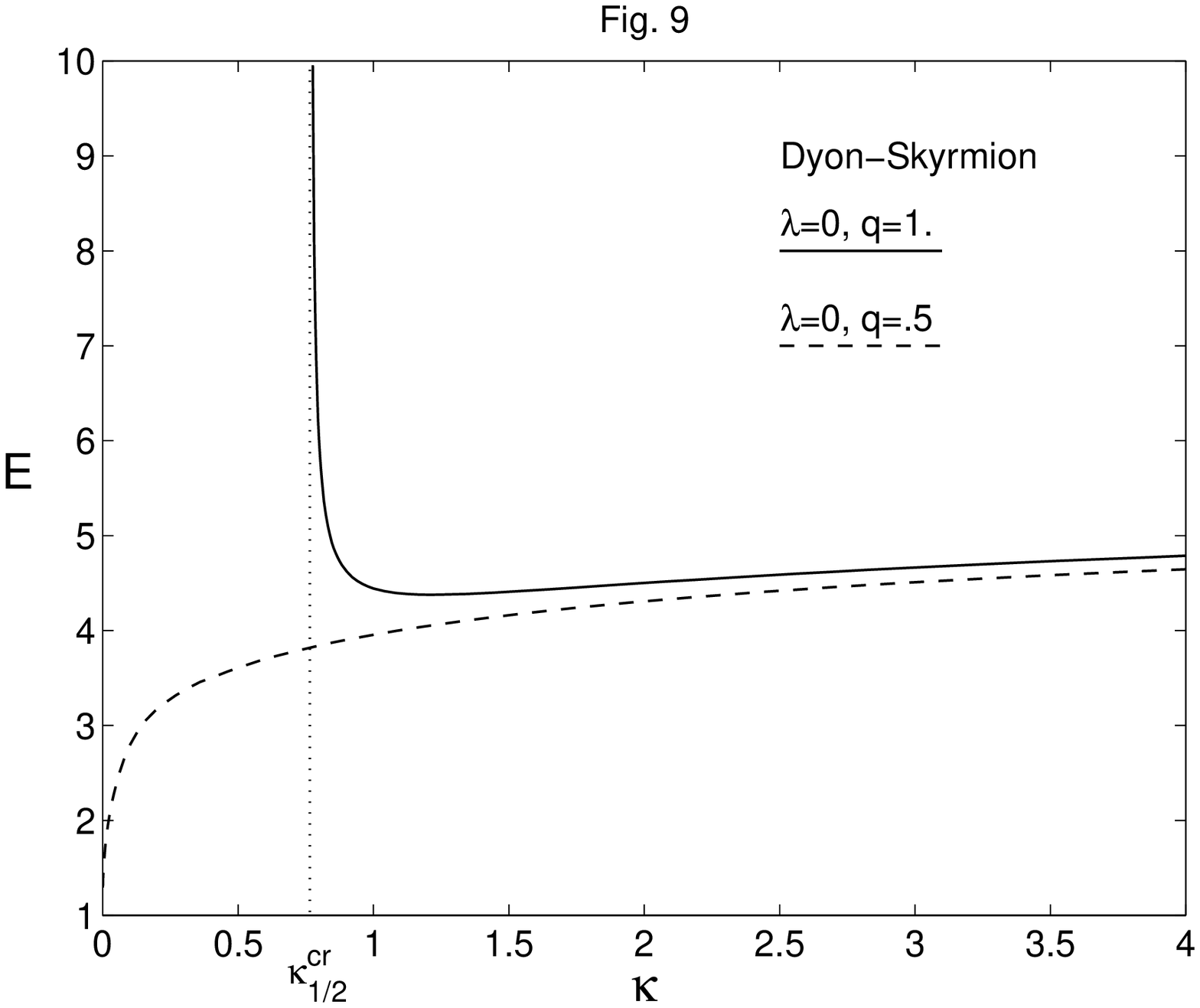}}
\end{figure}

\end{document}